\documentclass[aps,prb, reprint, showkeys, nofootinbib, oneside, superscriptaddress, longbibliography]{revtex4-2}
\usepackage{amsmath}
\usepackage{amssymb}
\usepackage{amsfonts}
\usepackage{mathtools}
\usepackage{physics}
\usepackage{xcolor}
\usepackage{color}
\usepackage{graphicx}
\usepackage{adjustbox}
\usepackage{placeins}
\usepackage[T1]{fontenc}
\usepackage{lipsum}
\usepackage{csquotes}
\usepackage[normalem]{ulem}
\usepackage[pdftex, pdftitle={Article}, pdfauthor={Author}]{hyperref} 
\usepackage{outlines}

\definecolor{dark-red}{rgb}{0.84,0.15,0.15}
\definecolor{dark-blue}{rgb}{0.15,0.15,0.8}
\definecolor{medium-blue}{rgb}{0,0,0.5}
\definecolor{copper}{rgb}{0.72, 0.45, 0.2}

\hypersetup{
    colorlinks, linkcolor={dark-red},
    citecolor={dark-blue}, urlcolor={medium-blue}
}
\usepackage[capitalize,nameinlink]{cleveref}

\def\equationautorefname~#1\null{Eq. (#1)\null}

\newcommand{\appref}[1]{\hyperref[#1]{App.~\ref*{#1}}}

\usepackage{wasysym}
\usepackage{enumitem}

\usepackage{ulem}
\makeatletter
\newsavebox{\@brx}
\newcommand{\llangle}[1][]{\savebox{\@brx}{\(\m@th{#1\langle}\)}%
  \mathopen{\copy\@brx\kern-0.5\wd\@brx\usebox{\@brx}}}
\newcommand{\rrangle}[1][]{\savebox{\@brx}{\(\m@th{#1\rangle}\)}%
  \mathclose{\copy\@brx\kern-0.5\wd\@brx\usebox{\@brx}}}
\makeatother

\begin{document}

\title{The source of hardware-tailored codes and coding phases}

\author{Gaurav Gyawali}
\email[Correspondence email address: ]{gaurav.gyawali@hpe.com}
\affiliation{Department of Physics, Cornell University, Ithaca, NY 14850, USA}
\affiliation{HPE Quantum Team, Hewlett Packard Labs, CA, USA}
\affiliation{Department of Physics, Harvard University, Cambridge, MA 02138, USA}

\author{Henry Shackleton}
\affiliation{Department of Physics, Massachusetts Institute of Technology, Cambridge, MA 02139, USA}
\affiliation{Department of Physics, Harvard University, Cambridge, MA 02138, USA}

\author{Zhu-Xi Luo}
\affiliation{Department of Physics, Harvard University, Cambridge, MA 02138, USA}
\affiliation{School of Physics, Georgia Institute of Technology, Atlanta, Georgia 30332, USA}

\author{Michael J. Lawler}
\email[Correspondence email address: ]{mlawler@binghamton.edu}
\affiliation{Department of Physics, Cornell University, Ithaca, NY 14850, USA}
\affiliation{Department of Physics, Applied Physics, and Astronomy, Binghamton University, Binghamton, NY 13902, USA}

\affiliation{Department of Physics, Harvard University, Cambridge, MA 02138, USA}

\date{\today} 

\begin{abstract}
A central challenge in quantum error correction is identifying powerful quantum codes tailored to specific hardware and determining their error thresholds above which quantum information is unprotected. This problem is hard because we cannot determine the noise models for our devices. Inspired by the quantum capacity theorem, we seek an optimal \emph{quantum source of information}, namely the density matrix that degrades minimally when passed through a noisy channel. 
We explore this idea with the Open Random Unitary Model (ORUM), a simplified model of a $N$-qubit quantum computer with competing depolarizing and dephasing channels as a stand-in for unitary gates and measurements. Through numerical optimization, we find that the ORUM hosts three discrete regimes, three ``phases'', the ``maximally mixed source'' phase, a ``$\mathbb{Z}_2$ source'' phase (where ORUM's $U(1)$ gauge symmetry is broken down to $\mathbb{Z}_2$), and a no-coding phase where all information is lost. These phases exhibit first-order transitions among themselves and converge at a novel \emph{zero-capacity multicritical point}.  These results show a remarkable similarity between the quantum capacity theorem and Jaynes' maximum entropy principle of statistical mechanics. Using the $\mathbb{Z}_2$ source, we build two codes, a classical cat code capable of correcting all the dephasing errors and a concatenated cat code capable of correcting all errors up to a distance $d=\text{min}(m,N)$ and reduces to Shor's 9-qubit code for $m=N=3$.  Neither classical nor quantum code survives near the vicinity of the zero-capacity multicritical point in the source phase diagram. Applying our approach to current noisy devices could provide a systematic method for constructing quantum codes for robust computation and communication.
\end{abstract}
\maketitle

Quantum error correction (QEC) faces a fundamental dichotomy. While theoretical efforts have yielded powerful generic QEC codes, such as asymptotically good qLDPC codes~\cite{panteleev_2021_asymptotically, bravyi_2014_homological}, these constructions often assume simplified noise models that ignore the complex, correlated realities of actual hardware. Conversely, experimental progress \cite{google2024suppressing, bluvstein2024logical, takeda2022quantum, sivak2023real}  has spurred a demand for hardware-tailored codes, such as the XZZX code, that exploit specific physical noise biases to maximize thresholds \cite{tuckett2018ultrahigh, bonilla2021xzzx,tuckett2019tailoring, dua_clifford_deformed_surface_codes_2024}. However, this tailoring process currently lacks a systematic, first-principles framework; it largely relies on heuristic adaptations of known codes or computationally expensive variational searches \cite{johnson2017qvector, xu2023_tailored_qec}.

A central challenge in designing hardware-tailored codes is that we do not have a deep understanding of the noise in our quantum devices. Improvements in gate fidelities often occur empirically, without a clear microscopic explanation for why the noise level has dropped~\cite{tuokkola2025methods}. Obtaining a rigorous description is practically intractable---standard quantum process tomography is exponentially hard, scaling as $16^n$ for $n$ qubits \cite{chuang_1997_prescription, poyatos_1997_complete}. Even modern techniques like shadow process tomography struggle in this context, suffering from measurement noise that complicates tomographic inversion and leads to a prohibitive bias-variance trade-off~\cite {hu2025demonstration}. Thus, characterizing the full noise profile remains a seemingly insurmountable challenge, and this appears to prevent the development of a solid theoretical foundation for hardware-tailored codes.

To circumvent this intractibility, a prevalent approach is to simplify the problem by assuming that the noise is ``biased'' in some way.  For example, if bit-flip errors are less likely than phase-flip errors we can tailor the surface code to be optimized for this specific noise structure \cite{tuckett2019tailoring}. In this single-qubit noise-biased setting, a new code was discovered to surpass the surface code, the XZZX code \cite{bonilla2021xzzx,xu2023tailored}, an example of Clifford-deformed surface codes \cite{dua_clifford_deformed_surface_codes_2024}. Beyond improved thresholds, these codes often yield practical benefits like efficient decoding and reduced overhead. Similarly, hardware platforms like Rydberg atom arrays can be engineered to convert dominant errors into erasures, a bias that improves surface code thresholds by a factor of four~\cite{wu2022erasure, scholl2023erasure, ma2023high}. However, these successes rely on the assumption of independent, local noise channels. Real-world devices suffer from complex, correlated failure modes---from vibrational modes in ion chains to cosmic-ray impacts and non-Markovian baths in superconducting circuits \cite{brownnutt2015ion, white2023filtering, harrington2025synchronous, bratrud2025measurement}---that defy simplified bias assumptions and require a more holistic optimization strategy.

Another strategy, given that we now have subthreshold quantum computers, is to take a data-centric approach to tailoring codes. We can infer the noise from the fidelity performance using machine learning \cite{nautrup2019optimizing}. We have seen some success with this approach. Reinforcement learning can discover codes that outperform the XZZX code under biased noise \cite{su2025discovery}, and transformer-based models can outperform leading surface-code decoders \cite{bausch2024learning}, demonstrating that noise has a learnable structure. Similarly, quantum neural networks such as quantum autoencoders have been used to discover hardware-tailored logical encoding \cite{locher_2023_qec_with_quantum_autoencoders}. We can also study noise syndromes and develop new decoding algorithms such as \verb!tesseract!, which improves coding by paying special attention to ancilla noise \cite{beni2025tesseract}. With strategies like these, we can learn empirically how to handle complex noise phenomena in our devices.

Finally, a principled framework needs to capture the collective phenomena arising in the competition between noise and coding. Error correcting thresholds are phase transition-like, for they require a thermodynamic limit to define precisely \cite{knill_1996_concatenated_quantum_codes, gottesman_1997_stabilizer_codes_qec, aliferis_2005_quantumaccuracy_threshold_concatenated, Chamberland_2016}. Additionally, stabilizer codes map to statistical mechanics and thus inherit the notions of phases, phase transitions, and diverging correlation lengths \cite{dennis2002topological, chubb_2021_stat_mech}. But this connection to statistical mechanics appears to extend beyond stabilizer codes. Decoupling theorems, which state that if the environment decouples, it has learned nothing, were built on the unitary model of black-holes, where a phase transition in the information available in the radiation, as observed in the Page curve, arises \cite{raju_2022_lessons, hayden2007black, hayden2008decoupling}. Quantum circuit models of measurement-induced phase transitions also exhibit a similar coding-no coding threshold \cite{li_2018_zeno, li_2019_measurement, skinner_2019_mipt, jian_2020_mipt, gullans_2020_dynamical, choi_2020_qec}. Thus, a theoretical foundation for hardware-tailored codes must do more than characterize local noise; it must account for collective, macroscopic phenomena of information persistence arising from the dynamic competition between code and noise.

We seek a guide to tailoring quantum codes to hardware by discovering the structure of quantum information that persists the longest. In quantum communication theory, as defined in the quantum capacity theorem, such a structure is known as an optimal \emph{quantum source} of quantum information. To explore this problem, we introduce the open random unitary model (ORUM), a simplified model of a quantum computer with two-qubit depolarizing channels arising from random unitary processes competing with one-qubit dephasing channels arising from weak measurements.  We find the optimal quantum source by numerically analyzing the ORUM's single-use quantum capacity, which we discover has a rich ``source phase'' diagram with two first-order transitions between distinct optimal sources, meeting at a continuous zero-capacity tricritical point. We use the language of statistical mechanics and the word ``phase'' because these numerical results suggest optimal quantum sources are discrete and the optimization problem is similar to Jaynes' maximum entropy principle. One source we identify is the $\mathbb{Z}_2$ source, after its relation to $\mathbb{Z}_2$ spin liquids and cat codes.  With the $\mathbb{Z}_2$ source as a guide, we construct two $\mathbb{Z}_2$ codes, one a classical cat code and one a quantum concatenated cat code, which, for a 3-qubit system concatenated 3 times, is Shor's 9-qubit code. In both cases, the threshold boundaries arise only deep into source phases, far away from the tri-critical point. We conclude by discussing how to determine the optimal quantum sources for physical devices and use them as a guide for the construction of hardware-tailored codes.

Quantum error correction is conventionally taught as follows. We begin with a logical state $|\psi\rangle_L = \sum_i \alpha_i |i\rangle_L$ in a logical Hilbert space $\mathcal{H}_L$ with basis $|i\rangle_L$. This state is encoded into a larger physical Hilbert space via an isometry $V : \mathcal{H}_L \to \mathcal{H}_{\mathrm{phys}}$, yielding the physical state $|\psi\rangle_{\mathrm{phys}} = V|\psi\rangle_L$. The goal is to construct $V$ such that most errors introduced by the environment take $|\psi\rangle_{\mathrm{phys}}$ out of the logical subspace (the Knill-Laflamme conditions \cite{knill1997theory}).

Quantum communication theorists view this same problem through the lens of correlations. Rather than tracking the isolated physical state, they track the information in the entanglement with a Reference $\mathcal{H}_R$ and the System $\mathcal{H}_S \equiv \mathcal{H}_{phys}$. They do so using a ``coherent copy'' isometry $W: \mathcal{H}_L \to \mathcal{H}_R \otimes \mathcal{H}_S$ that maps the logical basis states to entangled pairs: $W|i_L\rangle = |i\rangle_R \otimes |i_L\rangle_S$. Hence the state they use is $W|\psi_L\rangle = |\Psi_{RS}\rangle = \sum_i \alpha_i |i\rangle_R \otimes |i\rangle_S$. The goal now is to decouple the environment and protect the entanglement \cite{hayden2008decoupling}, i.e., protect the correlations from environmental noise. 

The two views are similar because they both must protect the system qubits from the same environmental noise to achieve their goals, and either view could be implemented in either physical setting. But they require different resources. In the remainder of the paper, we will adopt the second view and use a reference system and multiple uses of a quantum channel to learn hardware-tailored codes.

\begin{figure}[t]
\includegraphics{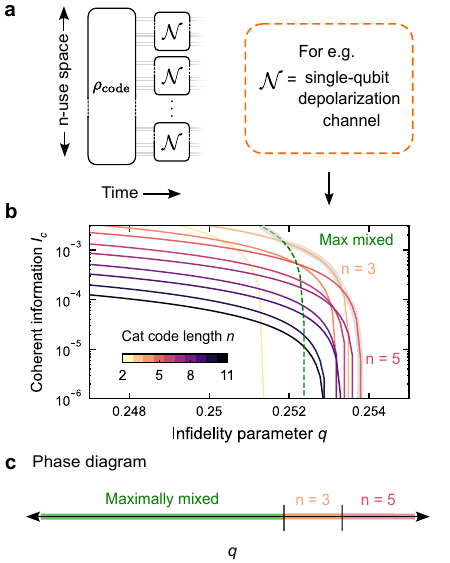}
    \caption{\textbf{Coding transition and superadditivity of the one-qubit depolarizing channel.} \textbf{a,} A general depiction of the many-body problem associated with many uses of a quantum channel $\mathcal{N}$. \textbf{b,} A reproduction of the discovery of superadditivity in the coherent information associated with the single qubit depolarizing channel \cite{divincenzo_1998_capacity_noisy_channels}.  While at low noise, the maximally mixed code wins, a 3-qubit cat code yields the highest coherent information $I_c$ at higher $q$, and a 5-qubit cat code wins at still higher $q$. No other cat codes competing out to 11-qubit cat codes. \textbf{c,} The coding phase diagram associated with discontinuities in the quantum capacity out to 10-uses as the depolarizing error rate $q$ is varied.}
    \label{fig:fig1}
\end{figure}

For a quantum channel $\mathcal{N}$, a completely positive trace-preserving map between density matrices, there is a maximum rate $Q^{(n)} = k/(nN)$ of sending $k$ qubits of information over $n$ ``uses'' of the $N$-qubit channel such that these qubits can be perfectly recovered on the other side \cite{lloyd_1997_capacity, shor_2002_quantum, devetak_2005_private_capacity}. This rate is computed using the \emph{coherent information} $I_c$ through the optimization problem \cite{schumacher_1996_dataprocessing, nielson_and_chuang}
\begin{equation}
\label{eq:channel_capacity}
Q=\lim_{n\to\infty}Q^{(n)} = \lim_{n\to\infty} \mathrm{max}_\rho I_c({\mathcal N}^{\otimes n},\rho)/(nN). 
\end{equation}
$I_c$ is most easily computed by purifying $\rho$ via $\rho = \Tr_R|\psi_{RS}\rangle\langle\psi_{RS}|$ using $R$, where it takes the form $I_c(\mathcal{N},\rho) = S_{RS}(\mathcal{N}(|\psi_{RS}\rangle\langle\psi_{RS}|) - S_S(\rho)$ with $S_A$ the von Neumann entropy of system $A$. 

The optimal density matrix $\rho_{opt}$ defines an optimal \emph{quantum information source}. Expressing it as an ensemble $\rho_{opt} \leftrightarrow \{p_i,|\psi_i\rangle\}_{i=1\ldots A}$ connects quantum sources to classical sources where we recognize the quantum states $|\psi_i\rangle$ are the letters of an alphabet, just like those used in Shannon's classical communication. Like classical sources play a fundamental role in classical communication and classical codes, quantum sources play a similar role in the quantum setting. 

\begin{figure*}[t]
\includegraphics{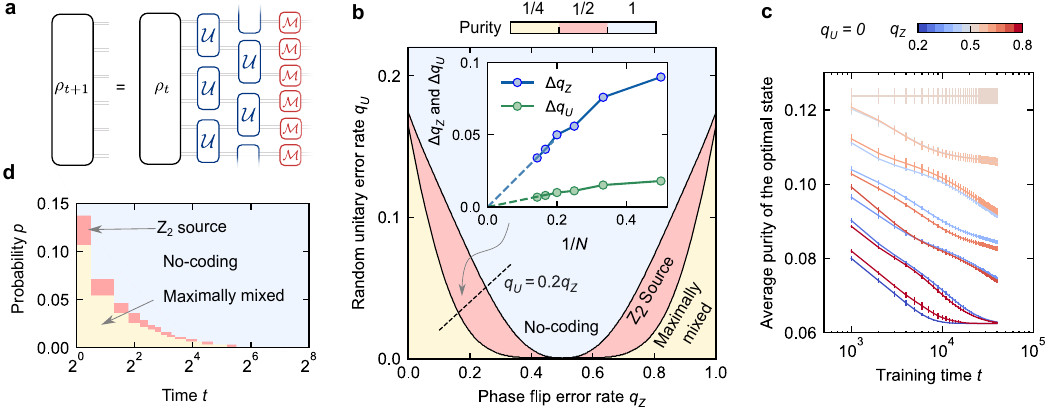}
    \caption{\textbf{Source phase diagram of open random unitary model (ORUM)} \textbf{a,} A quantum channel-based master equation for ORUM. \textbf{b,} Single-use/source phase diagram of the ORUM for a single time step showing the maximally mixed code phase, $\mathbb{Z}_2$ source phase, and no-coding regions. The inset shows the width of the $\mathbb{Z}_2$ source phase decays like $1/N$ with the system size $N$ along the dashed black line given by $q_U = 0.2 q_Z$. $\Delta q_U$ and $\Delta q_Z$ are horizontal and vertical components of the line connecting the maximally-mixed source region to the no-coding region along $q_U = 0.2q_Z$. \textbf{c,} Average purity of the optimal state found by the gradient-descent based optimizer showing divergence in optimization time close to the critical point $q_{Zc}=0.5$. \textbf{d,} Dynamical phase diagram of the ORUM obtained by taking $p = 2q_Z/q_U$.}
    \label{fig:fig2}
\end{figure*}

It is helpful to consider a simple example. Fig. \ref{fig:fig1} explores the optimization problem associated with the single qubit depolarizing channel $\mathcal{N}(\rho) \equiv {\mathcal U}_1(\rho) = (1-q)\rho + q \mathbb{I}/d$, where $\rho$ is any $N=1$ qubit state, and $d=2$, and we introduce the notation $\mathcal{U}_1$, $\mathcal{U}_2$, $\ldots$, for a 1, 2, or larger depolarizing channel arising from averaging $\rho$ over Haar random unitaries. Following Ref. \cite{divincenzo_1998_capacity_noisy_channels}, we first define a family of $n$-qubit ``cat sources'',
\begin{equation}
    \rho_{\mathrm{cat}}^{n} = \frac{1}{2}\left(|000\ldots\rangle\langle 000\ldots| + |111\ldots\rangle\langle 111\ldots|\right).
\end{equation}
We then send each of their qubits, one at a time, through ${\mathcal U}_1$ with error rate $q$, and compute the coherent information for this process. We call these states ``cat sources'' since their is an ensemble decomposition $\{(p_0,|000\ldots\rangle), (p_1,|111\ldots\rangle)\}$ that can be used to construct a cat code. Among the 10 such states presented, the 5-qubit cat source has the highest coherent information and is therefore the optimal source, given we are allowed up to 10 uses of the channel. The surprise that such multiple uses of a channel can enable more information to be sent through it per qubit is known as the superadditivity property \cite{elkouss2015superadditivity,leditzky2023generic} of the coherent information \cite{wilde_2013_qit}. It implies that the optimal state for an infinite number of uses, i.e. $\lim_{n\to\infty}Q^{(n)}$, is a challenging problem even for the one-qubit depolarizing channel, and it turns out, in this case, the optimal state allowing for any number of uses is still unknown \cite{wilde_2013_qit, shor_1996_qec_need_not, divincenzo_1998_capacity_noisy_channels, kianvash_2022_bounding}. From the vantage point of this finite size calculation, we can compute $Q^{(10)}$ by using $\rho_{cat}^{5}$ twice as the optimal source if $q\lesssim 0.252$, where $q$ is close to the threshold value. At lower values of $q$, the value of $Q$ goes through a \emph{first order source phase transition} where  $\rho_{cat}^{5}$ stops being the best source and $\rho_{cat}^{3}$ takes over. At even lower error rates, the maximally mixed source, which can be decomposed into an ensemble over any basis, finally wins. Hence, by optimizing the coherent information, we can identify optimal sources of quantum information that are robust to small changes in noise levels and can undergo source phase transitions.

It is less clear from the example how to construct quantum codes. We will illustrate this in our next example. 


Consider now the open random unitary model (ORUM), consisting of a brickwork layering of two-qubit depolarizing channels with error rate $q_U$ and one qubit dephasing channels with error rate $q_z$, schematically depicted by the master equation in \cref{fig:fig2}a. Depolarizing channels can arise from averaging over unitaries. In this model we choose the 2-qubit depolarizing channels to be
\begin{equation}
\mathcal{U}_{ij}(\rho) = (1-q_U)\rho + \int dU_{ij} U_{ij}\rho U_{ij}^\dagger
\end{equation}
where $U_{ij}$ is a 2-qubit unitary acting on qubits $i$ and $j$ and drawn from the Haar measure. Here $q_U$ is the rate at which we draw random unitaries, the rate at which we depolarize. We take the one-qubit dephasing channels to be
\begin{equation}
\mathcal{M}_i(\rho) = (1-q_z)\rho + q_z Z_i\rho Z_i.
\end{equation}
where $Z_i$ is the Pauli-Z operator acting on qubit $i$. At the special point where the phase error rate $q_z = 1/2$, $\mathcal{M}_i$ is equivalent to averaging over measurements. These channels preserve the $U(1)$ gauge symmetry,

\begin{align}
\mathcal{N}\left(e^{i\theta_iZ_i}\, \rho \, e^{-i\theta_iZ_i} \right) = e^{i\theta_iZ_i}\mathcal{N} \,(\rho) \, e^{-i\theta_iZ_i},
\end{align}
for $\mathcal{N} \in \{ \mathcal{U}_{ij}, \mathcal{M}_{i}\}$, as discussed in supplementary information (SI) \cref{sec:numerical_optimization}. So the ORUM is a toy model of a quantum computer, is built from commonly studied quantum channels, and has a high degree of symmetry. 

In \cref{fig:fig2}b, we plot the purity of the optimal state $\rho_{opt}$ obtained by numerical optimization on small system sizes up to $N=7$, as discussed in SI \cref{sec:numerical_optimization}. The purity reveals a remarkable discreteness in the resulting $\rho_{opt}$ sources. At low noise, the maximally mixed source dominates, while at high noise, a purity 2 source dominates before a no-coding regime with vanishing $Q$ emerges as expected from the theoretical bound requiring its presence above $q_U = 0.5$~\cite{roofeh_2024_phase_transition_qcapacity}. Remarkably, we find a discreteness in our solutions, though error rates change continuously, the optimal source remains the same. For this reason, we call \cref{fig:fig2}b a \emph{source phase diagram}. 

The purity 2 source has the form
\begin{equation}
\rho_{\mathbb{Z}_2} = \frac{1}{2}\left(|\mathrm{even}\rangle\langle \mathrm{even}| + |\mathrm{odd}\rangle\langle \mathrm{odd}|\right).
\end{equation}
It is like $\rho^n_{cat}$ but where the two states $|\mathrm{even}\rangle$, and $|\mathrm{odd}\rangle$, are more complex due to \emph{breaking ORUM's $U(1)$ gauge symmetry}. These two states are an equal magnitude random phase superposition of all computational basis states with an even/odd number of qubits in their $|1\rangle$ state. In the gauge where all phases are zero, it reduces to a cat source in the $X$-basis. We discuss more details of the $\mathbb{Z}_2$ source state in SI \cref{sec:z2_spin_liquid}. While the notion of a phase is apparent in the discreteness of the numerical solutions, this breaking of ORUM's local $U(1)$ symmetry down a global $\mathbb{Z}_2$ symmetry demands a ``phase transition'' between either the maximally mixed source phase or the no-coding phase and is our strongest evidence that source phases exist much like phases in statistical mechanics. 

As the inset shows, the $\mathbb{Z}_2$ source occupies a vanishing region of the phase diagram in the $N\to\infty$ thermodynamic limit, which means it becomes a sub-optimal source to the maximally mixed source, similar to a meta-stable phase, where, if it were practical to do so, we would use the maximally mixed source as our basis for building codes, over the $\mathbb{Z}_2$ source. However, here, we are interested in hardware-tailored codes and so to learn more about them and how one might use quantum sources to build them, we will pursue the construction of $\mathbb{Z}_2$ codes and the problem of how to do so using the $\mathbb{Z}_2$ source as a guide.

The highlight of the whole source phase diagram is the zero capacity continuous critical point at $q_U = 0$, $q_z = 0.5$. The vanishing channel capacity at $q_Z = 0.5$ can be inferred from the Hashing bound \cite{wilde_2013_qit}. At this point, vastly more states become competitive. Any pure state on the system qubits, for example, is in competition, as it has zero coherent information. We find analytically that $Q^{(1)}$ is infinitely differentiable along the entire line $q_U = 0$ (see SI \cref{sec:analytical_study_of_tricritical_point}). We further show, in \cref{fig:fig2}c, that the optimization time near this critical point diverges exponentially, demonstrating the existence of a complex set of suboptimal states near this point. This is a numerical demonstration of the change in competition arising near the critical point. Thus, the ORUM has a complex source phase diagram with two source phases and a novel \emph{zero-capacity continuous critical point} signified not by a singularity in $Q^{(1)}$ but by it vanishing at just one point.

A positive $Q$ value is an achievable rate of sending quantum information through one time step and recovering it perfectly on the other side. This is the content of the constructive proofs of the quantum capacity theorem \cite{lloyd_1997_capacity,shor_2002_quantum,devetak_2005_private_capacity,wilde_2013_qit}.  Repeating the encoding and recovery for $t$ time steps tells us we can preserve quantum information for any time $t$. However, the resources needed to do so are high. Any other error correction code than that used in the quantum capacity proofs will produce subdominant results. In what follows, we will build such subdominant but practical codes using the $\mathbb{Z}_2$ source aiming to preserve information for as long as possible and illustrate how biased noise can be exploited through the discovery of a biased source.

\begin{figure}[t]
    \includegraphics[width=0.475\textwidth]{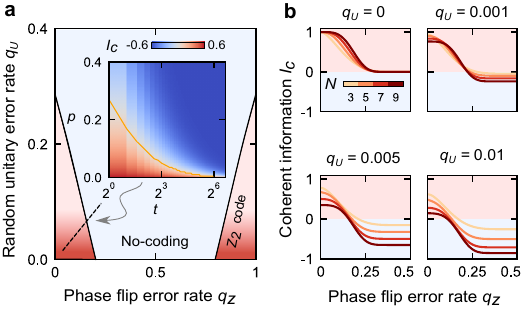}
    \caption{ \textbf{Classical $\mathbb{Z}_2$ Coding phase diagram.} \textbf{a}, Phase diagram of the classical $\mathbb{Z}_2$ code for a 3-qubit random unitary model (ORUM) with periodic quantum error correction (QEC) applied. At long times, the classical $\mathbb{Z}_2$ Code vanishes from the $q_U>0$ region of the phase diagram due to $X$-errors it cannot correct, here depicted by the fading red color. It survives at $q_U=0$ for $q_z \lesssim 0.25$ and $q_z\gtrsim 0.75$ where no $X$-errors arise. The inset shows the computed dynamics of coherent information along $p=2q_Z/q_U$, showing the phase vanishing in constant time along a line cut. \textbf{b}, Coherent information of the classical $\mathbb{Z}_2$ code, scanning along the $q_Z$ axis with $q_U$ fixed at $0$, $0.01$, $0.005$, and $0.01$ respectively for various system sizes $N$ at $t=N$. Finite-size scaling of the $q_U=0$ (top left) plot suggests a second-order phase transition around $q_z=0.25$ (and similarly for $q_z = 0.75$ not shown).}
    \label{fig:fig3}
\end{figure}

\begin{figure}[b]
    \includegraphics{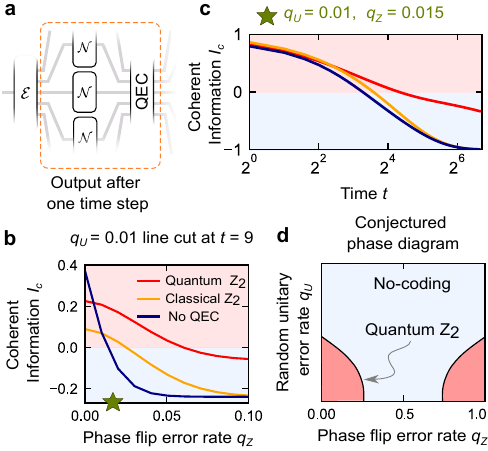}
    \caption{\textbf{Perfect communication via many uses of a noisy channel} \textbf{a}, Output of passing a $N$-qubit code through $m$-uses of a noisy channel $\mathcal{N}$ and a QEC layer after an encoding step $\mathcal{E}$. \textbf{b}, Coherent information $I_c$ of a 3-qubit $\mathbb{Z}_2$ code for $3$-uses of the ORUM channel i.e., $(N,m) = (3,3)$. $I_c$ are shown for $\mathbb{Z}_2$ code with no-QEC  (blue), classical $\mathbb{Z}_2$ code with QEC , and the quantum $\mathbb{Z}_2$ code with QEC and repetition $t=9$ (right) for $q_Z$ line cuts along $q_U=0.01$. Quantum $\mathbb{Z}_2$ has a clear advantage over classical $\mathbb{Z}_2$ code. \textbf{c}, Dynamics of the same three cases presented in \textbf{b} at $q_U=0.01, q_Z=0.015$, indicated by the green star, showing improvement in the communication rate at long times by leveraging the bit-flip and phase flip correction. \textbf{d}, A conjectured cartoon phase diagram of the quantum $\mathbb{Z}_2$ code in the thermodynamic limit i.e., $m,N,t \rightarrow \infty $ based on the existence of a QEC threshold for concatenated distance-3 codes \cite{aliferis_2005_quantumaccuracy_threshold_concatenated}.}
    \label{fig:fig4}
\end{figure}

Beginning with the $\mathbb{Z}_2$ source in the gauge with all phases set to zero, we find an ensemble decomposition $\rho_{\mathbb{Z}_2}\to\{(p_+,|+++\ldots\rangle),(p_-,|---\ldots\rangle)\}$ by working directly in the $X$-basis. This ensemble alone describes the classical $[N,1,N]$ repetition code, when $N$ is odd, and can correct up to $(N-1)/2$ simultaneous phase-flip errors. The error syndromes are known to be the parity checks $X_jX_{j+1}$ for $j \in [0, N-1]$. Correcting phase-flip errors using syndrome extraction and error correction every time step produces a drastic change to the source phase diagram, producing the \emph{coding phase diagram} as presented in \cref{fig:fig3}. Due to the absence of bit-flip errors at $q_U=0$, the infinitely differentiable zero-capacity critical point splits into coding-no-coding second-order phase transitions at $(q_z,q_U) \approx (0.5\pm 0.25,0)$, and the no-coding region grows to fill all $q_U>0$ due to the presence of uncorrectable $X$-errors generated by the depolarizing channel. The situation is, therefore, analogous to $q_U$ playing the role of temperature with coding phases existing at $q_U=0$ but not at any finite $q_U$. Though a qualitative change to the source phase diagram of \cref{fig:fig2}, this coding phase diagram is realizable in modern quantum devices . 

To improve upon this, what we need is a second classical code that can be combined with the above cat code to produce a quantum code. The natural choice here is to follow Shannon \cite{wilde_2013_qit}. Viewing the decompostion $\rho_{\mathbb{Z}_2}\to\{(p_+,|+++\ldots\rangle),(p_-,|---\ldots\rangle)\}$ as defining an alphabet, such as by letting $A\to |+++\ldots\rangle$ and $B\to |---\ldots\rangle$, we can produce \emph{codewords} and coded \emph{messages} by concatenating the letters. An example is the $m=3$ message $|+++\ldots\rangle\otimes|---\ldots\rangle\otimes|+++\ldots\rangle$ where it is understood that each ``letter'' is sent separately though the channel. Superpositions of such messages enable phase flip error detection. If the letters are selected according to their frequencies in the ensemble, then each use of the channel corresponds to sending $\rho_{opt}$ through it. In this way, we could construct random codes.

A simple approach is to utilize concatenation, similar to the first threshold theorems \cite{knill_1996_concatenated_quantum_codes,gottesman_1997_stabilizer_codes_qec}. Here, we again employ repetition but will return to the $|\mathrm{even}\rangle$ and $|\mathrm{odd}\rangle$ basis. If we replace the $|\mathrm{even}\rangle$ letter with multiple copies of it $|\mathrm{even}\rangle^{\otimes m}$ and similarly for $|\mathrm{odd}\rangle$, then a bit-flip error, which sends $|\mathrm{even}\rangle$ to $|\mathrm{odd}\rangle$ and vice versa, can be detected as it would move us out of the code space. Here, $m$ denotes the level of repetition. All that is left to do, then, is find a set of stabilizers that simultaneously stablize these two states and we will have defined a \emph{quantum} $\mathbb{Z}_2$ code.

The original stabilizer code for each subsystem gives us stabilizers $X_iX_{i+1}$ that correct phase flip errors. In addition, the repetition introduces the additional stabilizers $Z^N_IZ^N_{I+1}$, with $Z^N_I$ the product of Pauli $Z$ operators on all $N$ qubits of subsystem $I$. These enable us to correct bit-flip errors. The resulting $[[mN,1,\mathrm{min}(m,N)]]$ quantum $\mathbb{Z}_2$ code at $N=3$, $m=3$, is Shor's 9-qubit code.
\cref{fig:fig4} presents numerical results demonstrating this code. Specifically, \cref{fig:fig4}b reveals that at non-zero depolarizing noise ($q_U=0.01$), the concatenated code exhibits positive coherent information, overcoming the bit-flip noise that destroys the information in the classical $\mathbb{Z}_2$ code. This robustness is also present in the dynamics in \cref{fig:fig4}c, where the coherent information for the quantum $\mathbb{Z}_2$ code decays more slowly than the classical counterpart. However, what we ultimately seek here is perfect communication over the noisy channel, for which we will make $m$ channel uses similarly to the derivation of quantum capacity theorem \cite{devetak_2005_private_capacity}. Furthermore, leveraging the threshold theorem for concatenated distance-3 codes \cite{aliferis_2005_quantumaccuracy_threshold_concatenated}, we conjecture that in the large-$m$, large-$N$ limit, this code corrects arbitrary $(\mathrm{min}(m,N)-1)/2$ errors, creating the finite dome-like region shown in \cref{fig:fig4}d.

\section*{Discussion}

In this manuscript, we presented a principled method for the construction of hardware-tailored codes. It is built on the quantum source, a density matrix that optimizes the coherent information. There are several natural questions we must address about our approach. How close is the connection between optimal quantum sources and statistical mechanics? What challenges are likely if we were to implement the proposed approach on physical devices? Finally, beyond repetition and random codes, what other methods are there to turn a quantum source into a competitive quantum code?

There are remarkably many analogies between the optimization problem defined by the quantum capacity theorem and Jaynes' maximum entropy principle, which we review for the quantum case in SI \cref{app:Jaynes}. Both optimize a version of entropy; both involve a version of a Hamiltonian, a linear operator in Hilbert space $\mathcal{H}$ or a linear operator in the space of positive semidefinite operators $\mathrm{PSD}(\mathcal{H})$. Similarly, both have an optimal density matrix; both have different kinds of ensembles, microcanonical and canonical in the statistical mechanics case and quantum computing and quantum communication in the coding case. Similarly, both have system-level observables, the Free energy in statistical mechanics and the quantum capacity in quantum information, and both have symmetry where optima can break this symmetry. For these reasons, we have borrowed language from statistical mechanics to describe the results presented.

The observation of symmetry breaking is especially important in the present context of hardware-tailored codes. It is when symmetry breaking occurs, in the form of the $\mathbb{Z}_2$ source that breaks the $U(1)$ gauge symmetry down to a global $\mathbb{Z}_2$, that we were able to take advantage of the noise to build better codes than a code designed to protect against generic noise. Symmetry breaking explains the value of hardware-tailored codes at the level of the collective effects of noise. 

There is at least one significant difference between statistical mechanics and the quantum capacity theorem.  There is no special value of free energy in statistical mechanics but there is a special value of the quantum capacity, namely zero. This special value played an important role in the continuous critical point observed above, where though the $Q$ was continuous and analytic in the vicinity of the critical point, it passes through zero where the structure of optimal states is substantially different. This difference gave rise to exponentially long optimization times which signal the arrival of the continuous critical point. 

We believe the connection between coding and statistical mechanics should enable us to leverage the long history of statistical mechanics to deepen our understanding of coding. Is it possible to use the renormalization group to discover hardware-tailored codes? Can one associate fixed points to our observed sources that explains the robustness to small changes in error rates? 
Given the exponential advantage in logical error rates from improvement in physical error rates, getting the most out of our hardware seems always beneficial and so we will likely want to exploit what we can from  this analogy with statistical mechanics to build hardware-tailored codes.

\begin{figure}[t]
    \includegraphics{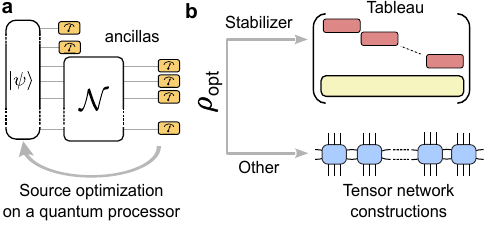}
    \caption{\textbf{Hardware-tailored code design} \textbf{(a)} Variational optimization of a source that maximizes the coherent information, \textbf{(b)} Taking an optimal code for a small system, we can construct more complex codes for practical application by appropriate concatenation of the tableau for stabilizer codes or by using tensor network constructions for other kinds of codes}.
    \label{fig:fig5}
\end{figure}

To tailor codes for modern quantum hardware, following the optimization approach presented here, a number of challenges need to be overcome.  A key step is estimating the von Neumann entropies that enter the computation of coherent information. The problem in general is challenging because the von Neumann entropy is a highly non-linear observable that is sensitive to noise. Fortunately, here it is the difference between the two large entropies that is needed. Polynomial algorithms are available for this situation \cite{audenaert2007sharp}. There are also other approximate techniques such as classical shadows \cite{vermersch2024enhanced}, a combination of matrix and phase estimation \cite{chowdhury2020variational}, and quantum neural estimation \cite{goldfeld2024quantum}. The feasibility of entropy measurement has been demonstrated by recent experimental results, including those on the Quantinuum and Google quantum processors \cite{greene2024measuring, gyawali_dfl}. In particular, Google's experiment measures the second Rényi entropy, a stand-in for the entanglement entropy, for up to 16 qubits. Likewise, we need long-lived reference qubits that stay idle after the initial state preparation, a resource enabled by tricks like twirling the noise and dynamical decoupling \cite{seif2024entanglement, Viola_dynamical_decoupling, das2021adapt}. We also ideally need multiple uncorrelated copies of the system by spatially separating the qubits into groups, each of which implements the channel, such as with multi-layer superconducting circuits \cite{matthews_2025_placing_and_routing} or with atom-based computers. Thus, applying our approach to tailoring codes on the quantum processors today is a promising future direction. 

Lastly, there is the question of scaling quantum sources up to hardware-tailored codes. While our investigation of the ORUM recovered a familiar family of concatenated codes, including the 9-qubit Shor's code, our framework facilitates a significantly broader class of hardware-tailored constructions. The optimal density matrix $\rho_{\text{opt}}$ serves as the fundamental ``brick'' we can use to build codes. So long as the state defined on $m$-copies reduces to this optimal source upon partial trace, the codewords will traverse the channel with minimal noise degradation. This principle underpins the workflow summarized in \cref{fig:fig5}. Starting with the discovered $\rho_{\text{opt}}$, if the resulting codewords form a stabilizer state, as was the case for the ORUM, we can scale up the code by concatenating stabilizer tableaus, patching bricks together as depicted in \cref{fig:fig5}b~, thereby leveraging standard error threshold theorems to guarantee robustness~\cite{knill_1996_concatenated_quantum_codes, gottesman_1997_stabilizer_codes_qec, aliferis_2005_quantumaccuracy_threshold_concatenated, Chamberland_2016}. In the more general case, tensor network frameworks like Quantum Lego allow us to connect the bricks together into a large-scale architecture \cite{quantum_lego_2022_cao}. Tensor networks are particularly well-suited for this task~\cite{tensor_networks_and_qec_2014_ferris, quantum_lego_2022_cao, pastawski_2015_holographic_QECC}, as they not only represent the complex code structure efficiently but also provide corresponding encoders and decoders~\cite{quantum_lego_2022_cao}. 
Crucially, this source-centric perspective extends to the frontier of error correction design. Whether serving as the seed for homological product constructions in asymptotically good qLDPC codes \cite{panteleev_2021_asymptotically, bravyi_2014_homological}, or defining the instantaneously protected subspace for dynamical Floquet codes \cite{hastings_haah_2021}, the discovered quantum source remains the fundamental primitive---a key to addressing the challenges of hardware-tailored quantum errror correction.

\section*{Methods}
\label{sec:methods}
To compute the single-use channel capacity $Q^{(1)}({\mathcal N})$, we define a trainable ``model'' using a dense representation (a Pytorch tensor) of $|\psi_{RS}\rangle$ as the initial state. Each element of the $|\psi_{RS}\rangle$ vector is a trainable parameter, much like weights in a neural network. Then we compute $I_c(|\psi_{RS}\rangle\langle\psi_{RS}|,{\mathcal N})$, where ${\mathcal N}$ is a collection of quantum channels that make up the whole quantum channel model. Each one-qubit and two-qubit channels are directly implemented through their corresponding superoperator, instantiated with non-trainable Pytorch tensors. The function ${\mathcal N}(|\psi_{RS}\rangle\langle\psi_{RS}|)$ itself was computed by combining these tensors using \verb!torch.einsum! to carry out the computations to produce the output density matrix $\rho_{R'S'}$. This density matrix is almost the largest tensor in memory, generally requiring $4n$ qubits worth of information for a $n$ qubit system $S$ (namely $\text{dim}(S)=n$, $\text{dim}(R)=n$, so its a $2^{2n}\times 2^{2n}$ matrix). The largest tensor in memory likely arises during the \verb!torch.einsum! call as an intermediary tensor. As such, slightly larger systems could be studied with a dense tensor representation with a more memory-efficient implementation of ${\mathcal N}$. Finally, we compute the von Neumann entropies $S(\rho_{R'S'})$ and $S(\rho_{S'})$ by straightforward diagonalization of the density matrices. The coherent information $I_c(|\psi_{RS}\rangle\langle\psi_{RS}|$, is then the ``loss function'' analogous to mean squared error typically employed in training neural networks, but with the intent to maximize it rather than minimize. Finally, we update $|\psi_{RS}\rangle$ using the gradients of the coherent information and normalize $|\psi_{RS}\rangle$ after each gradient ascent iteration.

\section*{Acknowledgement}
We thank Cenke Xu, Sarang Gopalakrishnan, Ehud Altman, Ashwin Vishwanath, Rahul Sahay, and Ruihua Fan for their valuable discussions during the early stages of this project.

\bibliography{main.bib}

\newpage
\onecolumngrid
\setcounter{equation}{0}
\setcounter{figure}{0}
\setcounter{table}{0}
\setcounter{section}{0}

\renewcommand{\thefigure}{S\arabic{figure}}

\makeatletter
\renewcommand{\thesection}{\arabic{section}}
\renewcommand{\thesubsection}{\thesection.\Alph{subsection}}
\renewcommand{\thesubsubsection}{\thesection.\alph{subsubsection}}
\renewcommand{\theequation}{S\arabic{equation}}
\renewcommand{\theHfigure}{S\arabic{figure}}
\renewcommand{\thetable}{S\arabic{table}}
\makeatother

\newpage
\begin{center}
    \textbf{Supplementary Information for\\[4mm]
\Large The source of hardware-tailored codes and coding phases} \\
    \vspace{5pt}
    Gaurav Gyawali, Henry Shackleton, Zhu-Xi Luo, Michael J. Lawler
\end{center}
\vspace{5pt}
\section{Numerical optimization of the channel capacity and the discovery of the \texorpdfstring{$\mathbb{Z}_2$}{Z2} source}
\label{sec:numerical_optimization}

\begin{figure}[b]
\begin{center}
\includegraphics[width=0.49\textwidth]{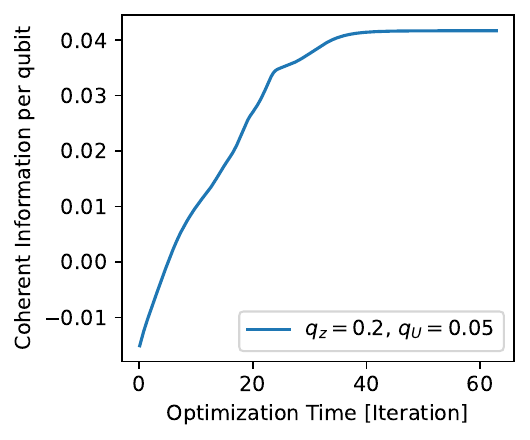}
\includegraphics[width=0.405\textwidth]{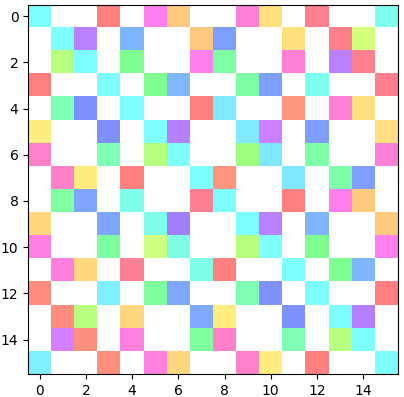}
\end{center}
\caption{\textbf{Discovery of the $\mathbb{Z}_2$ source for $N=2$.} {\bf a} optimization trace showing the rise of the coherent information $I_c$ as a function of the gradient ascent optimization step. {\bf b} An image plot of the ($16 \times 16$) density matrix, corresponding to 2 system and 2 reference qubits,  elements with white corresponding to zero matrix elements and color corresponding to matrix elements with a magnitude of 1/16 and phase given by the color. The light blue matrix elements have zero phase, the light red matrix elements have a phase of $\pi$ while all other colors are some other values of the phase between 0 and $2\pi$.}\label{fig:Z2discovery}
\end{figure}
 
The result of the optimization procedure described in the Methods is a density matrix $\rho_{opt}$ which is a local maximum of the coherent information. For the point $q_z = 0.1$, $q_U = 0.05$ in the ORUM phase diagram, an optimization trace is shown in Fig. \ref{fig:Z2discovery}a showing smooth monotonic trajectory with a few small bumps along the way that presumably correspond to other nearby local maxima or saddlepoints. We then plot the density matrix $\rho_{opt}$ in \cref{fig:Z2discovery}b as an image, finding that each matrix element has either zero magnitude or a magnitude of $1/2^N$, where $N$ is the number of qubits in the system. Each non-zero matrix element has a phase which is 0 along the diagonal and generally non-zero in the off-diagonal elements. A quick computation of the purity reveals $\Tr \rho_{opt}^2=1/2$. By converting the two and column indices of the non-zero elements to binary, we find that they either correspond to two numbers with an even number of ones in the binary strings or an odd number of ones. We go into further detail in \cref{sec:z2_spin_liquid}. In this way, optimization has revealed optimal input density matrices that represent encoded information with a surprising simplicity.

We numerically optimize the coherent information over all possible input density matrices to the ORUM channels. By using PyTorch \cite{pytorch} to implement the density matrix and channel superoperators, we optimize using autograd techniques familiar in machine learning together with a careful attention to preserving the norm of the density matrix. In the Methods section, we present details on how this optimization was carried out and on how we analyzed the resulting optimal density matrices to discover the $\mathbb{Z}_2$ source. Similarly, in \cref{fig:Z2discovery}, we present the optimization trace as well as the optimal density matrix that corresponds to the $\mathbb{Z}_2$ source.

\section{\texorpdfstring{$\mathbb{Z}_2$}{Z2} source as a spin liquid}
\label{sec:z2_spin_liquid}
The $\mathbb{Z}_2$ code optimal density matrix, found in section \ref{sec:numerical_optimization}, is defined as follows. Let $|\mathrm{even}\rangle_S$ be the state
\begin{equation}
    |\mathrm{even}\rangle_S = \frac{1}{2^{(n-1)/2}}\big(|00\ldots000\rangle_S + e^{i\phi_1}|00\ldots011\rangle_S + e^{i\phi_2}|00\ldots101\rangle_S + \ldots\big)
    \label{eq:even}
\end{equation}
where the sum extends over all basis states of the system $|x\rangle_S$ with an even number of 1's in the bit string $x$, each state having an arbitrary phase $\phi_i$. Then define the state $|\mathrm{odd}\rangle_S$ to be the state
\begin{equation}
    |\mathrm{odd}\rangle_S = \frac{1}{2^{(n-1)/2}}\big(|00\ldots001\rangle_S + e^{i\phi_3}|00\ldots010\rangle_S + e^{i\phi_4}|00\ldots100\rangle_S + \ldots\big),
    \label{eq:odd}
\end{equation}
where the sum extends over all basis states of the system $|x\rangle_S$ with an odd number of bit strings, again each state having an arbitrary phase. Then, we can define a purification of this state observed in the quantum capacity calculations to be
\begin{equation}\label{eq:z2_code_rs}
     \ket{\psi_{\mathbb{Z}_2}}_{RS}=\frac{1}{\sqrt{2}}\left(|0\rangle_R\otimes|\mathrm{even}\rangle_S + |1\rangle_R\otimes|\mathrm{odd}\rangle_S\right).
\end{equation}
and tracing over $R$ shows us this state is
\begin{equation}
    \label{eq:z2_code_s}
    \rho_{\mathbb{Z}_2, S} = \frac{1}{2}(|\mathrm{even}\rangle_S {}_S\langle \mathrm{even}| + |\mathrm{odd}\rangle_S {}_S\langle \mathrm{odd}|)
\end{equation}

The arbitrary phases that enter the purification suggest the existence of a gauge symmetry. Applying the unitary transformation $U(\{\theta_i\}) = e^{i\sum_j\theta_jn_j}$ with $n_j=(1+ Z_j)/2$ just alters the phases $\phi_1$, $\phi_2$, etc. in $|\mathrm{even}\rangle_S$. Applying a similar transformation but with $n_1 = (1-Z_1)/2$, $n_j = (1+Z_j)/2$, $j\geq 2$ to $|\mathrm{odd}\rangle_S$ similarly alters its phases. But if we set $\theta_i = \pi$, then we find $|\mathrm{even}\rangle_S \to |\mathrm{even}\rangle_S$ and $|\mathrm{odd}\rangle_S \to -|\mathrm{odd}\rangle_S$ which leaves $\rho_{\mathbb{Z}_2,S}$ invariant. 

To show ORUM has a gauge symmetry, let us first define it more carefully. We can write the two-qubit depolarizing channel as
\begin{equation}
{\mathcal U}_{i,j}(\rho) = (1-q)\rho + q\sum_{a,b}\tau_{ia}\tau_{jb}\rho\tau_{ia}\tau_{jb}
\end{equation}
where $\tau_{ia}$ are the identity and three Pauli operators $I_i$, $X_i$, $Y_i$, $Z_i$ on site $i$ indexed by $a$. We can also write the one-qubit dephasing channel as
\begin{equation}
{\mathcal Z}_i(\rho) = (1-q)\rho + qZ_i\rho Z_i
\end{equation}
Then, if we pass a gauge-transformed state $U(\{\theta_j\})\rho U^\dagger(\{\theta_j\})$ through the ORUM quantum channel, we find
\begin{equation}
    \mathcal{Z}_i\left(U(\{\theta_j\})\rho U^\dagger(\{\theta_j\}\right) = U(\{\theta_j\})\mathcal{Z}_i(\rho) U^\dagger(\{\theta_j\})
\end{equation}
and
\begin{equation}
    \mathcal{U}_{i,i+1}\left(U(\{\theta_j\})\rho U^\dagger(\{\theta_j\}\right) = U(\{\theta_j\})\mathcal{U}_{i,i+1}(\rho) U^\dagger(\{\theta_j\}
\end{equation}
for $U(\{\theta_j\})\tau_{ia}\tau_{i+1,b}U^\dagger(\{\theta_j\}) = \tau'_{ia}\tau'_{i+1,b}$ amounts to rotating the Pauli operators to a new local basis which still preserves the one-design property of these unitaries. Namely, after the gauge transformation, our expression for ${\mathcal U}_{i,i+1}$ has a different set of Kraus operators which define the same channel (Krauss operators are not unique). Hence, the phases we find numerically in $|\mathrm{even}\rangle_S$ and $|\mathrm{odd}\rangle_S$ are a consequence of a $U(1)^{\otimes n}$ gauge symmetry.  

As a result of the gauge symmetry, the $\mathbb{Z}_2$ source is like a $\mathbb{Z}_2$ spin liquid, it breaks the $U(1)$ gauge symmetry of the dynamics down to the global $\mathbb{Z}_2$ symmetry of $\rho_{\mathbb{Z}_2,S}$.

\section{Similarities between quantum capacity and the maximum entropy principle}
\label{app:Jaynes}

In 1957, Jaynes, inspired by Shannon's invention of information theory \cite{shannon_1948}, rederived statistical mechanics from an information perspective via the maximum entropy principle \cite{jaynes1957information,jaynes1957information2}. This principle states that the entropy is maximized in equilibrium, subject to constraints of what is known about a system. Here we revisit this principle for quantum statistical mechanics to draw attention of the similarity between it and the quantum capacity theorem discussed in the main text. 

The extension from classical statistical mechanics to quantum statistical mechanics was likely recognized immediately after Jaynes's papers, and a good early discussion on the topic is present in Douglas J. Scalopino's thesis \cite{scalapino1961irreversible}. One modern accessible reference is John Preskill's notes on quantum Shannon theory \cite{preskill2022}, where the Gibbs state is derived from a free energy minimization principle, but the focus of this treatment is on information theory, not statistical mechanics. 

The maximum entropy principle states that the entropy should be maximized subject to known constraints. For a closed statistical mechanics system, the von Neumann entropy, viewed as a function of the density matrix, is constrained by the known energy $E = \Tr \rho H$ of the system fixed by the initial conditions. We want to maximize the entropy with respect to varying $\rho$ while maintaining the energy constraint. But, since $\rho$ itself is not a simple matrix but one defined by additional constraints,  we need to view it as a general matrix subject to the additional unit trace $\Tr\rho=1$, hermiticity $\rho^\dagger = \rho$, and non-negativity $\rho\geq0$ constraints. Since hermiticity can be handled by imposing it directly on $\rho$ and non-negativity can be checked after an optimum is found, this just amounts to the requirement of imposing a unit trace. Hence, we can proceed by imposing just two constraints using the method of Lagrange multipliers. 

We begin with the functional:
\begin{equation}
    S = -k_B\Tr\rho\log\rho - \alpha(\Tr\rho - 1) -k_B\beta(\Tr\rho H - E) 
\end{equation}
When $\rho$ obeys both constraints, this functional is the entropy. A natural solution is Boltzmann's equal a priori probabilities, the mixed state $\rho$ of all states with $\Tr\rho H = E$, i.e.
\begin{equation}
    \rho(E) = \frac{1}{\Omega}\delta_\Delta(H - E)
\end{equation}
where $\Omega = \Tr\delta_\Delta(H-E)$ and $\delta_\Delta$ is a regularized $delta$-function of the Hamiltonian matrix $H$ minus $E$ with an energy width or cutoff $\Delta$.

An alternative solution, which avoids a singular distribution, is to change variables from energy $E$ to temperature $T$ via a Legendre transformation
\begin{equation}
F = E - TS = \Tr\rho H + k_BT\Tr\rho\log\rho + T\alpha(\Tr\rho - 1),
\end{equation}
where we set the Lagrange multiplier $\beta = 1/k_B T$. Taking the derivative with respect to complex variable $\rho_{ij}$, with $\rho_{ji} = \rho_{ij}^*$ treated as a separate variable, we obtain
\begin{equation}
    \frac{\partial F}{\partial\rho_{ij}} = \Tr E_{ij} H + k_BT\Tr E_{ij}\log\rho + k_BT\Tr\rho\frac{\partial}{\partial\rho_{ij}}\log\rho + T\alpha\Tr E_{ij}
\end{equation}
where $E_{ij} = \partial\rho/\partial\rho_{ij}$ is the matrix with matrix element $ij$ equal to one, i.e. $(E_{ij})_{ij} = 1$, and all others zero. To make sense of this expression, we need to work out the derivative of $\log\rho$. 

Defining the log of $\rho$ using a Taylor series
\begin{equation}
\log\rho = \log(I + (\rho-I)) = \rho-I -\frac{1}{2}(\rho-I)^2 + \frac{1}{3}(\rho-I)^3 + \ldots,
\end{equation}
the derivative is 
\begin{multline}
\frac{\partial}{\partial\rho_{ij}}\log\rho = E_{ij} -\frac{1}{2}\left(E_{ij}(\rho-I)+(\rho-I)E_{ij}\right) +
\frac{1}{3}\left(E_{ij}(\rho-I)^2 + (\rho-I)E_{ij}(\rho-I) + (\rho-I)^2E_{ij}\right) + \ldots.
\end{multline}
Multiplying this expression by $\rho$, taking the trace, and exploiting the cyclic property of the trace, we arrive at
\begin{equation}
\Tr\rho\frac{\partial}{\partial\rho_{ij}}\log\rho = \Tr\rho E_{ij}\left(I-(\rho-I)+(\rho-I)^2-\ldots\right).
\end{equation}
Recognizing this as a geometric series we see it simplifies to
\begin{equation}
\Tr\rho\frac{\partial}{\partial\rho_{ij}}\log\rho = \Tr \rho E_{ij}\frac{I}{I+(\rho-I)} = \Tr\rho E_{ij}\rho^{-1} = \Tr E_{ij}
\end{equation}
Hence, we obtain
\begin{equation}
    \frac{\partial F}{\partial\rho_{ij}} = \Tr E_{ij} H + k_BT\Tr E_{ij}\log\rho + (k_B+\alpha)T\Tr E_{ij}
\end{equation}
Now performing the trace leaves us with the matrix equation
\begin{equation}
\frac{\partial F}{\partial\rho_{ij}} = H_{ji} + k_BT(\log\rho)_{ji} + (k_B+ \alpha)T\delta_{ij} = 0
\end{equation}
Or
\begin{equation}
H + k_BT\log\rho +(k_B+\alpha)TI = 0
\end{equation}
with solution
$\rho = \frac{1}{Z}e^{-\beta H}$
with $Z = e^{-(k_B+\alpha)T}$ chosen to satisfy $\Tr\rho=1$. In this way, we see that the maximum entropy principle is a derivation of statistical mechanics from a quantum information theory principle.

\section{Measurement-induced phase transition}\label{sec:mipt}
\begin{figure*}[th]
\includegraphics{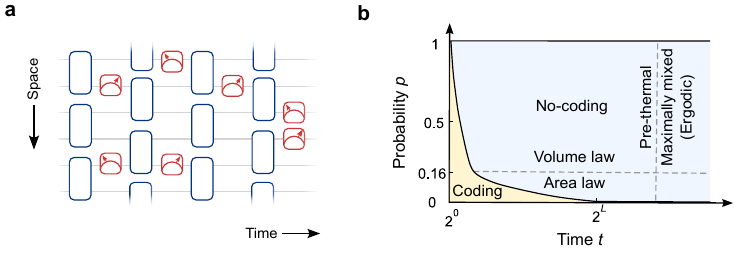}
    \caption{\textbf{Purification in monitored circuits} \textbf{a,} Four layers of a monitored circuit (top), commonly used to study measurement-induced phase transition (MIPT), consisting of random 2-qubit unitaries (blue rectangles) and interspersed measurement gates (red squares) inserted with some measurement probability $p$. \textbf{b,} The purification phase diagram of a random Clifford model with the maximally mixed initial state. Below the critical point ($p<p_c$), the purification time diverges with the system size, whereas, for $p>p_c$, the maximally mixed state purifies at constant time.}
    \label{fig:figS1}
\end{figure*}

The viewpoint of coding transition has been valuable to understanding measurement-induced phase transitions (MIPT) \cite{gullans_2020_dynamical,choi_2020_qec}. MIPT arises due to a dynamical interplay between unitary gates vs. projective measurements, resulting in an entanglement phase transition from a volume law to an area law scaling of the entanglement entropy \cite{li_2018_zeno, li_2019_measurement, skinner_2019_mipt, jian_2020_mipt}. Furthermore, these transitions are also reflected in the divergence of purification time for a mixed initial state. 
This is a coding transition, as the purification signifies that the system no longer carries information about the initially entangled reference \cite{gullans_2020_dynamical, choi_2020_qec}. While all states ultimately purify under monitored dynamics, they can reliably transmit quantum information for exponentially long timescales relative to system size. This phenomenon is captured in the system's coherent information, a key measure of its capacity to carry quantum information \cite{schumacher_1996_dataprocessing}. While the coding transition provides an alternative perspective to MIPT, observing it experimentally is still challenging due to the exponential sampling complexity of post-selection \cite{noel_2022_mipt, koh_2023_mipt_ibm, hoke_2022_google_mipt}. Nevertheless, identifying coding transitions by viewing the measurement-driven circuit as a noisy quantum channel generically doesn't require post-selection.

A typical circuit used to observe MIPT is shown in \cref{fig:figS1}a (top), which consists of a ``brickwork'' circuit of two-qubit unitaries from, for instance, random Clifford ensemble interspersed with single-qubit measurements in the $Z$-basis. Starting from a pure state, the probability of measurement $p$, continuously drives the steady-state entanglement entropy from a volume-law to an area-law scaling \cite{li_2018_zeno, li_2019_measurement, skinner_2019_mipt}. However, one can also start from a mixed state, for instance, the maximally mixed state, resulting from an initial entanglement of the system $S$ with a reference $R$. Although for any $p>0$, the system purifies at exponentially long times, the purification time exhibits two distinct behaviors. Above a critical measurement probability $p_c$, the purification time is constant, whereas it sharply diverges with respect to the system size for $p>p_c$ (bottom). The $p_c$ for the purification phase transition has been found to be the same as $p_c$  for MIPT \cite{gullans_2020_dynamical}. The purification transition is a ``coding transition'' because the system transitions from transmitting finite quantum information via the maximally mixed state to transmitting zero quantum information. The coherent information $I_c=S(\rho_{S}') - S(\rho_{RS})$ reflects the ability of an initial state $\rho_{RS}$ to transmit quantum information through a channel $\mathcal{E}$.

Our work examines the coherent information by passing various states through the channel $\mathcal{E}$. The maximum coherent information defines the channel capacity, see \cref{eq:channel_capacity}.  The trajectories generated by the quantum channel, which is often expressed in an operator-sum representation with Kraus operators, are quantitatively different from those produced by the original channel. A given channel can have many Kraus representations, so the procedure for generating trajectories is not unique if we take the channel as the fundamental dynamical law. For instance, a $N$-qubit depolarization channel can be defined as a sum over either Pauli, Clifford or the Haar-random unitaries as long as they form a 1-design. At a trajectory level, the entanglement phase transition for the Haar-random case has a higher $p_c$ compared to the Clifford case. However, there is no transition in the entanglement phase in the Pauli case. 

\section{Analytical study of the tricritical point}

\label{sec:analytical_study_of_tricritical_point}
\begin{figure*}[t]
\includegraphics{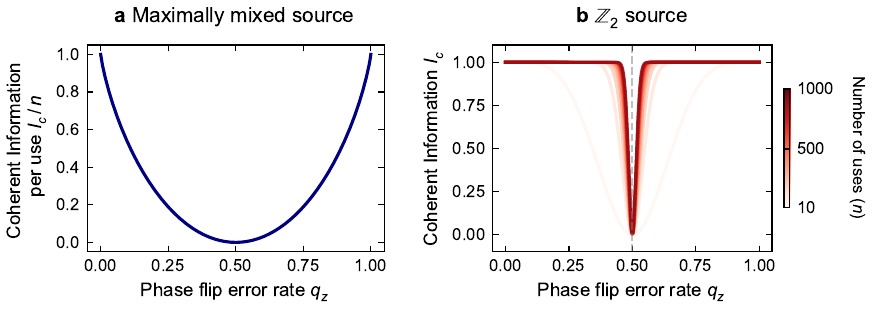}
    \caption{\textbf{Coherent information along $q_U=0$.} Coherent information of \textbf{a}, the maximally mixed code and \textbf{b}, the $\mathbb{Z}_2$ source along the line $q_U=0$. Fort both the codes $I_c$ vanishes at the multicritical point ($q_Z,q_U)=(0.5, 0)$.}
    \label{fig:figS2}
\end{figure*}  

In this section, we analytically study the coherent information for the two types of sources in the coding regions near the tricritical point located at $(q_Z, q_U) = (1/2, 0)$. By restricting our analysis to the axis $q_U=0$, the Open Random Unitary Model (ORUM) simplifies to a pure dephasing channel, rendering the spectrum of the output density matrices analytically tractable. We explicitly derive the coherent information $I_c(q_Z)$ for both the symmetry-respecting maximally-mixed source and the symmetry-breaking $\mathbb{Z}_2$ source. Explicit derivation of coherent information via statistical mechanics mapping for more sophisticated QEC codes subject to various other noise models can be found in Refs \cite{colmenarez_2025_fundamental_thresholds,colmenarez_2024_accurate, huang_2025_coherent_information}. Our derivation demonstrates that both quantities vanish at the critical threshold $q_Z=1/2$ and, crucially, that the channel capacity remains infinitely differentiable along this trajectory, showing the continuous nature of the zero-capacity critical point observed in the numerical optimization.

\subsection{Coherent Information of the maximally mixed code along $q_U=0$}
Consider a single qubit in the maximally mixed state $\rho_1 = \mathbb{I}/2$. We purify this state with a reference qubit $R$ into the Bell state in the $X$-basis:
\begin{equation}
    \ket{\Phi^+}_{RS} = \frac{1}{\sqrt{2}}\left( \ket{+}_R\ket{+}_S + \ket{-}_R\ket{-}_S \right).
\end{equation}
The single-qubit dephasing channel $\mathcal{M}$ applies the Pauli-$Z$ operator with probability $p = q/2$. In the $X$-basis, the $Z$ operator acts as a bit-flip: $Z\ket{+} = \ket{-}$ and $Z\ket{-} = \ket{+}$. The action of the channel on the system qubit maps the initial state $\ket{\Phi^+}_{RS}$ to the orthogonal state $\ket{\Psi^+}_{RS}$:
\begin{equation}
    (I \otimes Z) \ket{\Phi^+}_{RS} = \frac{1}{\sqrt{2}}\left( \ket{+}_R\ket{-}_S + \ket{-}_R\ket{+}_S \right) = \ket{\Psi^+}_{RS}.
\end{equation}
Thus, the output joint state is a classical mixture of these two orthogonal Bell states:
\begin{equation}
    \rho'_{RS} = \left(1-q_Z\right) \ket{\Phi^+}\bra{\Phi^+} + q_Z \ket{\Psi^+}\bra{\Psi^+}.
\end{equation}
Since $\ket{\Phi^+}$ and $\ket{\Psi^+}$ are orthogonal, the von Neumann entropy is simply the Shannon entropy of the mixing probabilities:
\begin{equation}
    S(\rho'_{RS}) = H\left(q_Z\right) = -q_Z\log_2q_Z - \left(1-q_Z\right)\log_2\left(1-q_Z\right).
\end{equation}
The reduced state of the system remains maximally mixed, $\rho'_S = \mathbb{I}/2$, with entropy $S(\rho'_S) = 1$. The single-qubit coherent information is therefore $1 - H(q_Z)$. For the full system of $n$ qubits, the result is:
\begin{equation}
    I_c(q_Z) = n \left[ 1 - H\left(q_Z\right) \right].
\end{equation}
We show the plot for $I_c/n$ in \cref{fig:figS2}a

\subsection{Coherent information of the $\mathbb{Z}_2$ source along $q_U=0$}
The coherent information of the ORUM channel for the $\mathbb{Z}_2$ source at $q_U=0$ can be computed analytically by determining the exact spectrum of the output density matrix. We consider the $\mathbb{Z}_2$ source state:
\begin{equation}
    \ket{\psi}_{\mathbb{Z}_2,RS} = \frac{1}{\sqrt{2}} \left( \ket{++\cdots +}_{RS} + \ket{-- \cdots -}_{RS} \right).
\end{equation}
Tracing out the reference system $R$ yields the mixed system state:
\begin{equation}
    \rho_{\mathbb{Z}_2,S} = \frac{1}{2} \left( \ket{++\cdots+}_S \bra{++\cdots+}_S + \ket{--\cdots-}_S \bra{--\cdots-}_S\right).
\end{equation}
Here, we consider the $q_U=0$ caase, so a code $\rho$ passes through $n$-uses of the single-qubit channel
\begin{equation}
    \rho' = \mathcal{M}(\rho) = \left(1-q_Z \right) \rho + q_Z Z\rho Z .
\end{equation}
After $n$-uses, we obtain
\begin{equation}
    \rho' = \mathcal{M}^{(n)}(\rho) = \sum_{k=0}^n \left(1-q_Z\right)^{n-k} \left(q_Z\right)^k \sum_{E_k} Z_{E_k} \rho Z_{E_k},
\end{equation}
where the inner sum runs over all $\binom{n}{k}$ error configurations $E_k$ consisting of Pauli-$Z$ operators acting on $k$ qubits. We define the probability weight for a weight-$k$ error as:
\begin{equation}
    \beta(k, q) = \left(1-q_Z\right)^{n-k} \left(q_Z\right)^{k}.
\end{equation}
To find the spectrum of $\rho'$, we analyze the action of the errors on the code basis states. Recall that the Pauli-$Z$ operator flips the $X$-basis states: $Z\ket{+} = \ket{-}$ and $Z\ket{-} = \ket{+}$. Let $\mathbf{x} \in \{+,-\}^n$ denote a basis string in the $X$-basis.
\begin{itemize}
    \item When a weight-$k$ error $Z_{E_k}$ acts on the term $\ket{+\cdots+}\bra{+\cdots+}$, it generates a state $\ket{\mathbf{x}_k}\bra{\mathbf{x}_k}$ where the string $\mathbf{x}_k$ has exactly $k$ minus signs.
    \item When a weight-$j$ error $Z_{E_j}$ acts on the term $\ket{-\cdots-}\bra{-\cdots-}$, it generates a state $\ket{\mathbf{x}'_j}\bra{\mathbf{x}'_j}$ where the string $\mathbf{x}'_j$ has $j$ plus signs (and thus $n-j$ minus signs).
\end{itemize}
The output density matrix $\rho'$ is diagonal in this $\{\ket{\mathbf{x}}\}$ basis. Consider a specific basis state $\ket{\mathbf{x}}$ containing $k$ minus signs. This state is populated by two sources:
\begin{enumerate}
    \item From the $\ket{+\cdots+}$ branch: via an error of weight $k$. This occurs with probability $\frac{1}{2}\beta(k, q)$.
    \item From the $\ket{-\cdots-}$ branch: via an error of weight $n-k$. This occurs with probability $\frac{1}{2}\beta(n-k, q_Z)$.
\end{enumerate}
Thus, the eigenvalues of $\rho'_{\mathbb{Z}_2,S}$ depend only on the Hamming weight $k$ of the basis state:
\begin{equation}
    \lambda_k = \frac{1}{2} \left( \beta(k, q_Z) + \beta(n-k, q_Z) \right).
\end{equation}
Since there are $\binom{n}{k}$ such states for each weight $k$, the von Neumann entropy is:
\begin{align}
    S(\rho'_{\mathbb{Z}_2,S}) &= - \sum_{k=0}^n \binom{n}{k} \lambda_k \log_2(\lambda_k) \nonumber \\
    &= - \sum_{k=0}^n \binom{n}{k} \frac{\beta(k, q_Z) + \beta(n-k, q_Z)}{2} \log_2\left( \frac{\beta(k, q_Z) + \beta(n-k, q_Z)}{2} \right).
\end{align}
Next, we calculate the joint entropy $S(\rho'_{\mathbb{Z}_2,RS})$. Since the reference + system states $\ket{+\cdots+}$ and $\ket{-\cdots-}$ are orthogonal, they effectively label the two branches. The channel action does not mix these branches in the joint basis. The joint density matrix is block diagonal, equivalent to a classical mixture of the two error distributions. The eigenvalues are simply $\frac{1}{2}\beta(k, q)$ (occurring twice for each $k$, once for each branch). However, since we sum over the full distribution which is normalized, this simplifies to the Shannon entropy of the error distribution:
\begin{align}
    S(\rho'_{\mathbb{Z}_2,RS}) &= - \sum_{k=0}^n \binom{n}{k} \beta(k, q_Z) \log_2(\beta(k, q_Z)) \equiv H(\beta).
\end{align}
Finally, the coherent information $I_c = S(\rho'_{\mathbb{Z}_2,S}) - S(\rho'_{\mathbb{Z}_2,RS})$ is given explicitly by:
\begin{equation}
I_c(q_Z) = \sum_{k=0}^n \binom{n}{k} \left[ \beta(k, q_Z) \log_2 \beta(k, q_Z) - \lambda_k \log_2 \lambda_k \right].
\end{equation} 
We plot this expression in \cref{fig:figS2}b, where we can see that $I_c$ for the $\mathbb{Z}_2$ source vanishes sharply at the critical point. In \cref{fig:figS2}a, we saw that the coherent information of the maximally mixed state vanishes smoothly at the critical point. Indeed, all three phases, namely, the $\mathbb{Z}_2$ source coding phase, the maximally-mixed coding phase, and the no-coding phase, all meet together at this critical point; hence, it is a tri-critical point. Furthermore, in the limit of infinitely many uses, the maximally-mixed code maximizes the coherent information infinitesimally close to the critical point, so the channel capacity is also given by \cref{fig:figS2}, an infinitely differentiable function.
\end{document}